\documentstyle [prl,aps,floats]{revtex}

\begin{document}

\title{Inflationary Universe in Higher Derivative Induced Gravity}

\author{ W.F. Kao\thanks{email:wfgore@cc.nctu.edu.tw}}
\address{Institute of Physics, Chiao Tung
University, Hsinchu, Taiwan}

\date\today
\maketitle

\begin{abstract}
In an induced-gravity model, the stability condition of an inflationary slow-rollover solution is shown to be $\phi_0 \partial_{\phi_0}V(\phi_0)=4V(\phi_0)$.
The presence of higher derivative terms will, however, act against the stability of this expanding solution unless further constraints on the field parameters are imposed. 
We find that these models will acquire a non-vanishing cosmological constant at the end of inflation.
Some models are analyzed for their implication to the early universe.
\end{abstract} \vskip .1in

PACS numbers: 98.80.Cq; 04.20 -q; 
\vspace{0.5cm}

In a scale-invariant model, all dimensionful parameters are functionals of the scalar field.  
Therefore, scale invariance provides a natural way resolving the physical origin of these dimensionful parameters.

Scale invariance is also known to be important in various branches of physics. 
For example, QCD \cite{ryder} and many other inflationary models have been studied in the literature \cite{zee,ni,acc,acc1}. 
Note that inflation resolves many problems of the standard big bang cosmology \cite{acc,acc1}.
These problems include the flatness, the monopole, and the horizon problem.
In addition, local scale (or Weyl) symmetry has been suggested to be related to the missing Higgs problem in electro-weak theory \cite{CK}. 
Weyl symmetry has also been the focus of many recent activities \cite{smolin,Lee}. 
Scale-invariant effective theory is also suggested to be important for the physics near fixed points of the renormalization group trajectory \cite{smolin}.

In addition, higher derivative terms should be important for the physics near the Planck scale \cite{kim,dm95}. 
For example, higher-order corrections derived from the quantum gravity or the string theory have been considered in the study of the inflationary universe \cite{green}.
Higher derivative terms also arise as quantum corrections to the matter fields \cite{green}.
Moreover, the stability analysis of the pure higher-derivative models was shown in Ref. \cite{dm95}. 
It is hence interesting to extend this stability analysis to different models. 
In an induced-gravity model, it turns out that stability conditions of an inflationary solution are that the scalar field must obey a set of scale-invariant conditions under the slow-rollover approximation. 
We will also study the implication of this constraint to the inflationary universe in this paper.

We will focus on the induced-gravity model with an $R^2$ coupling \cite{wang} given by
\begin{equation}
  S=\int d^4x \sqrt{g} \left\{ -{ 1 \over 2 }  \epsilon {\phi}^2 R
   - { 1 \over 2 } g^{\mu\nu} {\partial}_{\mu}{\phi} {\partial}_{\nu}{\phi}
   - V(\phi) -{\alpha \over 3}  R^2 \right\}
\label{action}
\end{equation}
in this paper. 
Here $\epsilon$ and $\alpha$ are dimensionless coupling constants.
$V(\phi)$ is any possible symmetry-breaking potential.
Note that there are additional fourth-derivative terms in the most general higher derivative theory. 
They are related to the $R^2$ term, due to the Euler constraint and the fact that the Weyl tensor vanishes in the Friedmann-Robertson-Walker (FRW) spaces \cite{dewitt}, in four-dimensional spaces.

We will define $sV_0 \equiv 4V_0-\phi_0 \partial_\phi V(\phi_0)$ with $s$ denoting the scaling factor, $\phi_0$ denoting the initial condition of the inflaton $\phi$, and $V_0 \equiv V(\phi_0)$. 
In addition, the case that $s=0$ will be referred to as the scaling condition in this paper.
Note that the scaling condition will be shown to be a direct consequence of the slow-rollover approximation. 
Hence the initial data has to be close to the scaling condition.
We are going to show, however, that the induced-gravity model tends to stabilize the inflationary phase.
This is true if initial conditions of the inflaton are close to the scaling condition.
The presence of the higher derivative term (HDT) will further impose strong constraints on field parameters and scalar potentials. 
These constraints are required to generate a stable inflationary phase.
Otherwise, this theory can not permit an exponentially expanding solution under the scaling condition in the presence of the HDT.

Note that our universe is homogeneous and isotropic to a very high degree of precision \cite{data}.
Such a universe is described by the well-known FRW metric \cite{weinberg1}. 
Therefore, we will work on the FRW metric that can be read off directly from the following equation:
\begin{equation}
ds^2 \equiv g_{\mu\nu} dx^\mu dx^\nu =  -dt^2 + {a^2}(t)
\Bigl( {dr^2 \over 1 - k
r^2} + r^2 d\Omega  \Bigr).
\label{eqn:frw} \label{FRW}
\end{equation}
Here $ d \Omega $ is the solid angle $d \Omega \, = \, d{\theta}^2 + {\sin}^2 \theta \, d{\chi}^2 $ and $k \, = \, 0, \pm 1$ stand for a flat, closed, and open universe respectively.


The Friedmann equation can be shown to be \cite{kao99}
\begin{eqnarray}
&& 3 \epsilon \phi^2 (H^2 + \frac{k}{a^2} + 2 H {\dot{\phi} \over \phi} )=
V+{1 \over 2} \dot{\phi}^2 +K .  \label{ih}
\end{eqnarray}
Here $K \equiv 
12 \alpha [ 2H \ddot{H} +6H^2 \dot{H} -\dot{H}^2
-2 H^2 {k / a^2} + {k^2 / a^4} ]$ denotes the contribution from the HDT.
Moreover, the Euler-Lagrange equation for $\phi$ is
\begin{equation}
\ddot{\phi} +3 H \dot{\phi} + {\partial V \over \partial \phi} =
6 \epsilon \phi ( \dot{H} + 2 H^2 + {k \over a^2} )    .   \label{ip}
\end{equation}
Note that the HDT does not affect the $\phi$ equation directly as shown above.

One will analyze the inflationary solution under the slow-rollover approximation such that $|\dot{\phi}/\phi|
 \ll H$, and $|\ddot{\phi}/\phi| \ll H^2 $ for a brief period of time.
The slow-rollover approximation will also be shown to be consistent with field equations.
Assuming that $\phi=\phi_0$ and $H=H_0 +\delta H$, one can perturb the Friedmann equation and the scalar field equation.
These perturbed equations can be employed to study the stability of the inflationary solution.
Accordingly, one can show that leading-order perturbation equations give
\begin{eqnarray}
3 \epsilon \phi_0^2 H_0^2&=& V_0  ,      \\
\phi_0 {\partial V \over \partial \phi} (\phi= \phi_0) &=&12 \epsilon \phi_0^2
H_0^2 .
\end{eqnarray}
Therefore, the initial data satisfies the scaling condition in this approximation.
In addition, linear-order perturbation equations give
\begin{eqnarray}
4 \alpha \delta \ddot{H} +12 \alpha H_0 \delta \dot{H} - \epsilon \phi_0^2
 \delta
H &=&0,  \label{leadingorder} \\
\delta \dot{H} + 4 H_0 \delta H &=&0. \label{leadingphi}
\end{eqnarray}
Therefore, one has
\begin{eqnarray}
\delta H &\sim& \exp (-4 H_0 t) , \label{con-low} \\
\epsilon \phi_0^2 &=& 16 \alpha H_0^2 \label{con-high} .
\end{eqnarray}
Eq. (\ref{leadingorder}) will not be present without the HDT. 
Therefore, one will not have the constraint (\ref{con-high}) accordingly.
Moreover, Eq. (\ref{con-low}) indicates that inflation tends to stabilize the inflationary phase under the scaling condition.
Note that Eq. (\ref{con-high}) is the extra constraint derived from the HDT. 
This indicates that the gravitational constant ($\epsilon \phi_0^2/2$) is related to the Hubble constant $H_0$ during the inflationary phase.
Therefore, a physically acceptable inflationary induced-gravity model will be affected significantly by the HDT.

In addition, the first-order perturbation equation shows that the inflationary solution is indeed stable against the perturbation  $\delta H$.
Therefore, inflation will remain effective for at least a brief moment while $\phi$ changes slowly. 
Note also that the $\phi$ equation states that 
$$\ddot{\phi} +3 H_0 \dot{\phi} \sim 0 $$ during the period when $H \sim H_0$.
This gives
\begin{equation}
\phi \sim \phi_0 + {\dot{\phi}_0 \over 3 H_0}[1 - \exp (- 3 H_0 t) ]. \label{phit}
\end{equation}
Therefore, the slow-rollover approximation is indeed consistent with field equations.
Consequently, if the initial data satisfies the scaling condition, the system will undergo a strong inflationary process and remain stable for a long period of time under the scaling condition.
Therefore, we will focus on the case that the initial data of the effective theory obeys the $s=0$ condition.

Note that leading-order perturbation equations give us a few constraints on the field parameters according to
\begin{equation} \label{constraint}
4V_0 =\phi_0 {\partial V \over \partial \phi} (\phi= \phi_0) =12 \epsilon
\phi_0^2 H_0^2=192 \alpha H_0^4 .
\end{equation}
This is equivalent to
\begin{eqnarray}
H_0^2&=& {\epsilon \phi_0^2 \over 16 \alpha }\label{h0eq}  , \\
4V_0 &=&\phi_0 {\partial V \over \partial \phi} (\phi= \phi_0)
={3 \epsilon^2 \over 16 \alpha} \phi_0^4 \label{consts} .
\end{eqnarray}
Therefore, there are indeed strong constraints on the possible form of the scalar field potential according to Eq. (\ref{consts}).
These constraints also relate the field parameters in a nontrivial way. 
We will come back to this point later and study the constraint equation for some extended $\phi^4$ models.

If the scaling condition is not obeyed closely, the inflationary solution will not be strictly stable. 
This will act in favor of the graceful-exit process.
In such cases, the scalar field will obey the following equation 
\begin{eqnarray}
\ddot{\phi} &+& 3 H \dot{\phi} + {\dot{\phi}^2 \over \phi}
+ \left[ \partial_\phi V - 4 V/ \phi \right]/(1 +6 \epsilon) \nonumber \\
 &=& \left[ \dot{K} + 4H K \right]/(1 +6 \epsilon)\phi  \label{ipo}
\end{eqnarray}
which can be derived from differentiating Eq. (\ref{ih}) and comparing it with Eq. (\ref{ip}). 
Note further that Eq. (\ref{ipo}) is equivalent to the $G_{ij}$ component of the Einstein equation. 
Even this equation is redundant, it is still very useful for our analysis.
In summary, the inflationary solution can not be stable unless (i) the scaling condition is closely obeyed and (ii) $\alpha$ is constrained by Eq. (\ref{con-high}).
In such cases, the dynamics of the scalar field can be depicted from Eq. (\ref{ipo}).
We are about to show that the case (ii) will be violated in the conventional $\phi^4$ SSB potential.
Therefore, the system will follow the evolutionary process similar to the one described in Ref. \cite{acc,acc1}.
On the other hand, the physics will be different when the initial data falls too close to the scaling condition. 

For comparison, the equation of motions will become
\begin{eqnarray}
&& (H^2 + \frac{k}{a^2} + 2 H {\dot{\phi} \over \phi} )=
V+{1 \over 2} \dot{\phi}^2 +K ,  \label{ihein} \\
&& \ddot{\phi} +3 H \dot{\phi} + {\partial V \over \partial \phi} = 0
\label{ipein}
\end{eqnarray}
for the $R^2$-corrected Einstein theory given by ${\cal L}= -R/2 - \alpha R^2 +{\cal L}_{\phi}$.
Hence the scaling constraint no longer holds here since the $\phi$-equation does not couple to the $R^2$ term directly. 
Indeed, one can show that $H_0^2=V_0/3$ from the zeroth-order perturbation equation.
Note that the perturbation is done with respect to $H=H_0+\delta H$ under the slow-rollover approximation $|\ddot{\phi}| \ll H |\dot{\phi}|$ and $H \gg |\dot{\phi}|$. 
One can also show that the first-order perturbation equation gives Eq. (\ref{leadingphi}) after setting $\epsilon \phi_0^2 \equiv 1$. 
Therefore, the effect of the $R^2$-corrected Einstein theory is different from the induced-gravity model.
Hence the scale-invariant initial condition is a very unique property of induced-gravity models.

For a physical application, we will consider the following effective symmetry-breaking potential
\begin{equation}
V={\lambda_1 \over 4} (\phi^2 - v^2)^2 + {\lambda_2 \over 4} \phi^4 -\Lambda.
\label{effV}
\end{equation}
We are about to show that the apparent cosmological-constant term $\Lambda$ has to be non-vanishing in order to admit a consistent inflationary solution.
In addition, it reduces to the standard $\phi^4$ SSB potential if $\lambda_2=\Lambda=0$.
Therefore, neither scale-invariant potential nor standard $\phi^4$ SSB potential can provide a physically acceptable inflationary solution under the influence of the HDT. 
Hence one has to introduce an alternative asymmetric potential in order to generate an inflationary solution.

Indeed, one can solve Eq. (\ref{consts}) and show that
\begin{equation}
\phi_0^2 = v^2 - {4 \Lambda \over \lambda_1 v^2}=
{16 \alpha\lambda_1 v^2 \over 16 \alpha (\lambda_1 +\lambda_2) - 3 \epsilon^2 }
\label{const1}.
\end{equation}
Therefore, one can derive
\begin{equation}
\Lambda = {\lambda_1 \over 4} v^4 \left[
{16 \alpha \lambda_2 - 3 \epsilon^2  \over
16 \alpha(\lambda_1 + \lambda_2) -  3 \epsilon^2 }
\right]. \label{Lambda}
\end{equation}
Writing $\lambda\equiv\lambda_1 + \lambda_2$, one can further show that the extended $\phi^4$ SSB potential reads
\begin{equation}
V= {\lambda \over 4} \phi^4
- ( {\lambda \over 2} - {3 \epsilon^2 \over 32 \alpha} ) \phi_0^2 \phi^2
+ ({\lambda  \over 4} - {3 \epsilon^2 \over 64 \alpha} ) \phi_0^4 \label{Veff}.
\end{equation}
This is the form of the most general extended $\phi^4$ SSB potential that could admit an inflationary solution. 
One can also show that the minimum of this potential is 
$V_m \equiv V(\phi_m)= ( 3 \epsilon^2  / 16 \alpha )
[ 1/4  - 3 \epsilon^2 /  64 \alpha \lambda  ]\phi_0^4$
when 
$\phi^2=\phi_m^2 \equiv (1-  3 \epsilon^2 /16 \alpha \lambda )\phi_0^2$.
In addition, one has
$V_m = ( 3 \epsilon^2 / 16 \alpha \lambda)V(0)$ 
where $V(0)\equiv V(\phi=0)$ is the maximum of $V$. 
Note also that $V_m < V(0)$ is consistent with the equation
$\phi^2=\phi_m^2 \equiv
(1- 3 \epsilon^2 / 16 \alpha \lambda )\phi_0^2$.
This implies that $3 \epsilon^2 / 16 \alpha \lambda <1$.
Hence one has $\alpha >0$ because that 
$V_0 =3 \epsilon \phi_0^2 H_0^2= 3 \epsilon^2 \phi_0^4/64 \alpha >0$. 
In addition, one expects $\lambda >0$ since $V'' (\phi_m) = 2 \lambda \phi_m^2 >0$ for a local minimum at $\phi_m$. 
Therefore, one shows that $V_m =3 \epsilon^2\phi_0^2 \phi_m^2/64 \alpha >0$. 

Moreover, the effective gravitational constant observed in the post-inflationary phase is related to $\phi_0$ by the identity
$1/4\pi G=\epsilon \phi_m^2=(1-3 \epsilon^2 /16 \alpha \lambda ) \epsilon \phi_0^2$.
In addition, the effective cosmological constant observed in the post-inflationary phase is 
$V_m = 3H_0^2/2$.
Here we have set $\epsilon \phi_m^2/2=1$ in Planck unit.
If the scale factor $a(t)$ is capable of expanding some $60$ e-fold
in a time interval of roughly $\Delta T \sim 10^8$ Planck unit, the Hubble constant should be of the order $H_0^2 \sim 10^{-6}$ in Planck unit.
Therefore, one ends up with a rather big cosmological constant of the order $10^{-6}$ if the extended $\phi^4$ model is in effect.

One can now show that the case $\Lambda=\lambda_2=0$ is problematic.
Indeed, $\Lambda=0$ implies that $ \lambda_2= 3 \epsilon^2 /16 \alpha$ from Eq. (\ref{Lambda}). 
Hence $\epsilon/\alpha=0$ if $\lambda_2=0$.
This is apparently inconsistent with our assumption that $\alpha$ is small and  $\epsilon$ is finite. 
In fact, the case that $\epsilon=0$ will lead to an infinite gravitational constant. Hence it should be ruled out. 
Therefore, the case that $\Lambda=\lambda_2=0$ can not support an inflationary phase if the HDT is present.

Note that the expansion rate $H_0= \sqrt{\epsilon \phi_0^2/16 \alpha}$ can be adjusted to accommodate $60$ e-fold expansion rather easily \cite{acc}. 
Note also that small $\alpha$, hence small higher-order correction, will act in favor of the inflationary process.
In addition, the slow-rollover approximation is taken care of automatically by the higher-order term. 
Therefore, the only constraint on ${\phi}_0$ is that it should be small according to Eq. (\ref{phit}).

Once the scalar field rolls down toward the minimum-potential state, inflation will come to an end.
In addition, the soft-expansion era in the post-inflationary phase will be dominated by another lower-order induced-gravity model \cite{smolin}. 
Therefore, the re-heating process will be taken over by that lower-order effective induced-gravity model \cite{acc}.
Hence the HDT, acting in favor of the inflation process, plays an important role during the inflationary phase. 
It is still true even if the higher-order correction is small, namely, $\alpha \ll 1$.
Note, however, that this model implies a non-vanishing cosmological constant.
This may have to do with the field contents of the early universe \cite{weinberg}.

In other words, the smallness of the cosmological constant is not resolved by this approach.
Something else has to help resolving the cosmological-constant problem.
One possibility already mentioned earlier is that this induced-gravity theory remains effective, only during the inflationary era, as an collective effect of the physics in the early universe \cite{smolin}. 
This effective induced theory will no longer be held responsible for the physics after inflation is completed.

In addition, one can also consider the following symmetry-breaking Coleman-Weinberg potential from radiative correction \cite{coleman}
\begin{equation}
V={\lambda_1 \over 4} \phi^4 \ln ({\phi \over  v})^4 + {\lambda_2 \over 4}
\phi^4 - \Lambda.
\label{effVc}
\end{equation}
Note that we will use the same notation for $V$, $\lambda_i$, $\Lambda$,
etc. for simplicity although we are working on a different model.
One can show that the first constraint in Eq. (\ref{consts}) gives
$\Lambda= -\lambda_1 \phi_0^4 /4$. 
The second one gives $V_0=3 \epsilon^2 \phi_0^4/ 64 \alpha$.
This implies that 
$\phi_0=v \exp [3\epsilon^2 /64 \alpha \lambda_1 - (\lambda_1 +
\lambda_2)/4\lambda_1]$. 
Therefore, one can put the potential as
\begin{equation}
V={\lambda_1 \over 4} \phi^4 \ln ({\phi \over  \phi_0})^4 + {3 \epsilon^2
\over  64 \alpha} \phi^4 - {\lambda_1 \over 4} (\phi^4-
\phi_0^4).
\label{effVcm}
\end{equation}

In addition, one can show that the minimum state occurs when 
$\phi=\phi_m = \phi_0 \exp [-3\epsilon^2 /64 \alpha \lambda_1]$.
Furthermore, one can show that
$V_m=(\lambda_1 \phi_0^4/4) \{ 1- \exp [-3\epsilon^2 /16 \alpha
\lambda_1] \}$.
Therefore, inflation can be achieved rather easily.
On the other hand, one can show that $V_m$ depends on the choice of the parameters $x\equiv \epsilon^2 /\lambda_1$ and 
$y\equiv 3/16 \alpha$ according to 
\begin{equation}
V_m = {e^{xy}-1  \over x}.
\label{vmxy}
\end{equation}
One can hence show that $V_m$ increases as $x$ decreases or $y$ increases.
This is proved by showing that $\partial_x V_m$ is always negative and $\partial_y V_m$ is always positive definite.
Hence by choosing smaller $\lambda_1$, larger $\alpha$, or larger $\epsilon$ would lead to a smaller $V_m$.
In practice, one should choose $xy \ll 1$ such that $V_m \to 3/16 \alpha$.
In addition, the condition $xy \ll 1$ is equivalent to the condition 
$\lambda_1 \gg 3 \epsilon^2 /16 \alpha = \epsilon^2 V_m (xy \ll 1)$.
Therefore, $\alpha$ has to be very large in order to push $V_m$ toward $0$.
This is somewhat inappropriate as $\alpha$, related to the particle contents during inflation, can be computed from their quantum corrections in curved space \cite{weinberg1}. 
Therefore, one expects $\alpha$ to be small.  
Hence it is not likely that one can tune field parameters in order to push $V_m$ to the limit of observation in this theory.
Note further that one has $\alpha >0$ because 
$V_0 =3 \epsilon \phi_0^2 H_0^2= 3 \epsilon^2 \phi_0^4/64 \alpha >0$. 
In addition, one expects 
$\lambda >0$ since $V'' (\phi_m) = 4 \lambda \phi_m^2 >0$ where $\phi_m$ is a local minimum. 
Hence one has $V_m>0$ if $\alpha \lambda_1>0$.

Note that the scaling condition is derived from the slow-rollover approximation. 
It was shown that the initial data has to be close to the scaling condition.
We have, however, shown that a stable inflationary solution exists only when (i) scaling condition is closely obeyed and (ii) $\alpha$ is constrained by Eq. (\ref{con-high}).
We also show explicitly that two different extended models are not able to produce a universe with a vanishing cosmological constant all alone.
Therefore, in the traditional $\phi^4$ SSB model under the scaling condition, inflationary solution does not favor a stable inflationary solution in the presence of the HDT. 
Accordingly, the traditional slow-rollover inflationary solution will soon fall off the scaling limit even if it started out close to the scaling condition. 
Hence, one does not need to worry about whether the scalar field will be frozen to the scaling condition. 
Therefore, the effect of the HDT will act in favor of the graceful-exit process. 

{\bf Acknowledgments :}
This work is supported in part by the National Science Council under
the contract number NSC88-2112-M009-001.

\end{document}